# RF-MEMS beam components: FEM modelling and experimental identification of pull-in in presence of residual stress

Alberto Ballestra[1], Eugenio Brusa[2], Giorgio De Pasquale[1], Mircea Gh. Munteanu[3], Aurelio Somà[1]

[1] Politecnico di Torino, Mechanical Department,
c.so Duca degli Abruzzi, 24 – 10129 Torino, Italy;
alberto.ballestra@polito.it, giorgio.depasquale@polito.it, aurelio.soma@polito.it,
phone ++39 011 5646951; fax ++39 011 5646999

[2,3] Università degli Studi di Udine, Dept. Electrical, Management and Mechanical Engineering
via delle Scienze 208 – 33100 Udine, Italy; [2] eugenio.brusa@uniud.it, [3] munteanu@uniud.it
phone [2]++39 0432 558299, [3]++39 0432 558243, fax ++39 0432 558251

*Abstract*-In this paper an experimental validation of numerical approaches aimed to predict the coupled behaviour of microbeams for out-of-plane bending tests is performed. This work completes a previous investigation concerning in plane microbeams bending.

Often out-of-plane microcantilevers and clamped-clamped microbeams suffer the presence of residual strain and stress, which affect the value of pull-in voltage. In case of microcantilever an accurate modelling includes the effect of the initial curvature due to microfabrication. In double clamped microbeams a pre-loading applied by tensile stress is considered. Geometrical nonlinearity caused by mechanical coupling between axial and flexural behaviour is detected and modelled.

Experimental results demonstrate a good agreement between FEM approaches proposed and tests. A fairly fast and accurate prediction of pull-in condition is performed, thus numerical models can be used to identify residual stress in microbridges by reverse analysis from the measured value of pull-in voltage.

## I. INTRODUCTION

Microcantilevers and microbridges are currently widely used in RF applications as microswitches and microresonators [1], [2] and in experimental micromechanics, where materials mechanical properties and strength are measured. Therefore, it is required implementing efficient numerical models to predict the electromechanical performance of microstructures actuated by electric field [2], [3], [4], [5]. A wide variety of approaches has been proposed in literature to predict static and dynamic behaviour of microbeams [1], [6], [7], [8]. Experimental validation is therefore aimed to verify their effectiveness in predicting pull-in condition and frequency response. Usually model sensitivity on the uncertainties of numerical values of design parameters and material properties is investigated. Very often is rather difficult to know precisely material properties and microspecimen dimensions.

FEM original approaches developed by the authors were proposed in [9], [10], [11], [12] and were already validated in [13]. A preliminary experimental investigation was aimed to predict the static behaviour of planar microcantilevers, for in-plane bending test. Present research is devoted to complete previous investigation activity focusing on out-of-plane bending microbeams.

Fully coupled electromechanical problem has electrical and mechanical coordinates, which are linked by the electromechanical coupling effect. Pull-in condition is responsible of a snap down of microbeam on the counter-electrode. In present case pull-in may be affected by some initial stress or strain present on the microsystem before the application of electric field. Moreover, it is well known that the problem is nonlinear because of the dependence of electromechanical force on displacement and voltage and, sometimes, of the so-called geometrical nonlinearity [10], [11], [12], [13]. To include these effects, models of microcantilever have to introduce the approximated analytical description of the initial curvature. It is usually sufficient to predict with enough accuracy the pull-in voltage and displacement, with an error of 2-3% maximum. For microbridges with double clamps axial stress is required to perform a coherent simulation of the actual system.

## II. SPECIMEN CHARACTERIZATION

### A. Fabrication process and measurement methods

Specimens used for this work were realized by ITC-IRST Research Center (Trento, Italy), by means of the so-called *RF Switch (RFS) Surface Micromachining* process. Gold is used for the suspended structures; material is deposited through electroplating by means of a chromium-gold PVD adhesion layer [14], [15]. Profilometric measures and pull-in tests were performed by *Fogale Zoomsurf 3D* optical profiling system, based on non-contact optical interferometry [16]. The lateral resolution is ±0.3 μm, while the vertical resolution reaches ±0.5·$10^{-4}$ μm [17].Tables 1 and 2 show the dimensions of microbeams used as specimens in testing, all measures are expressed in micron. In particular, $t$ means beam thickness, $L$ is length, $w$ is width, $g$ initial gap, $y$ tip displacement for cantilever. Star symbol indicates measured value instead of nominal. Pull-in voltage ($V_{PI}$) was measured by gradually increasing voltage between suspended microbeam and ground, until the collapse of the structure on the ground counter-electrode.

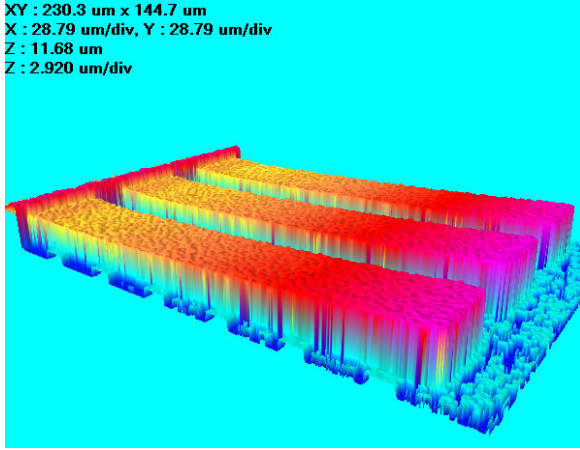

Fig. 1-Geometry 4 series cantilever beams (L=200 um)

Experimental pull-in was observed between the voltage range reported in tables I e II.

*B. Residual stress*

Presence of an initial curvature in microcantilever specimens is associated to an increase of pull-in condition. In double clamped microbeams pull-in depends on a pre-stress tensile loading caused by microfabrication process. Residual stress origin is supposed to be diffusion of chrome of adhesion layer among gold grains [15] and difference of thermal expansion coefficient (CTE) between the gold structure and the photoresist layer underneath [18]. Residual stress grows every time there is a temperature difference during microfabrication process. Stress is maximum at the interface gold/chromium/photoresist and decreases accross the microbeam thickness. In microcantilever, once removed the support substrate, traction at the interface is removed and turns into a deformation. Therefore initial curvature can be seen as an initial strain, which bends the microstructure out-of-plane [17]. In microbridges free bending is forbidden since microbeam is overconstrained by clamps. Therefore residual stress holds and strain is prevented .

### III. Modelling

According to the detailed descriptions of numerical models available in [1], [2], [3], [4], [5], [6], this paper will focus on the experimental validation of numerical results computed. Few modelling issues will be here resumed. Electro-mechanical force couples mechanical (displacements, rotations) and electrical (voltages, charges) degrees of freedom, and equilibrium equations in both the static and dynamic domains appear nonlinear. Moreover, typical dimensions of RF-MEMS introduce a second cause of nonlinearity, being referred to as "geometrical". In case of microcantilever a large displacement of the tip requires to resort to a nonlinear mechanical solution to find the actual equilibrium condition. This usually means to implement an iterative procedure which applies force step by step and locally linearizes the structural problem. In case of double clamped microbeam even in presence of small displacement the mechanical coupling occurring between the axial and flexural behaviour introduces either a hardening or softening effect on the structural stiffness. In both cases above mentioned, to solve the coupled problem a sequential solution is performed. Computation of the electric field distribution and of the related electromechanical force for a given equilibrium configuration of the deflected microbeam is separated from the mechanical analysis aimed to find the deformed shape of the beam under the electromechanical load. This sequence justifies implementing a computational loop. If the geometrical nonlinearity is active, a second iterative loop has to be implemented for each step of the electromechanical solution to find the actual equilibrium condition. Numerical methods are used to discretize the structure and the dielectric material. In present case both the dielectric and mechanical domains were meshed by FEM. A sequential approach based on Newton Raphson iteration method was implemented. The coupled electromechanical problem was solved by 2D and 3D models, by means of ANSYS code. In case of dynamic analysis Newmark Modified Algorithm was implemented and tested in connection with the non incremental approach for geometrically nonlinear structures. ANSYS code, MATLAB and FORTRAN implementations were performed and numerical results were compared to the experimental ones.

Alternately the coupled-field problem was solved by using coupled-field elements through a direct approach available in ANSYS as 1D transducer element TRANS126. It couples electro-mechanical domains and consists of a reduced-order model with structural translations and electric potential as degrees of freedom.

Initial deformed shape of microcantilver was analytically described by means of beam curvature $\kappa$ and axial strain $\varepsilon$ being written as function of flexural displacement $v$, axial displacement $u$, axial load $N$ and bending moment $M$ as follows:

$$\kappa = -\frac{d^2v}{dx^2} = \frac{M}{EJ}; \qquad \varepsilon = \frac{du}{dx} = \frac{N}{EA} \qquad (1)$$

Generalized force vector $F$ was then formulated as:

$$\{F\} = \{F\} - \sum_{i=1}^{n_{el}} \left( \int_l [B]^T \begin{Bmatrix} M_0 \\ N_0 \end{Bmatrix} dx \right) + \sum_{i=1}^{n_{el}} \left( \int_l [B]^T [D] \begin{Bmatrix} \varepsilon_{0T} \\ \kappa_{0T} \end{Bmatrix} dx \right) \qquad (2)$$

where initial values of force $N_0$ and Moment $M_0$ can be computed from the initial stress $\sigma_0$; $\varepsilon_{0T}$ is the initial thermal strain and $\kappa_{0T}$ the curvature in case of thermal contribution.

TABLE 1
CANTILEVER BEAMS: NOMINAL AND MEASURED (*) DIMENSIONS, MEASURED AND CALCULATED PULL-IN VOLTAGES

| Geom. | Sample | t | L | L* | w* | t* | g* | y* | $V_{PI}$* Exper. | $V_{PI}$ Trans 126 | $V_{PI}$ Ansys 2D | $V_{PI}$ Matlab 2D | $V_{PI}$ Ansys 3D |
|---|---|---|---|---|---|---|---|---|---|---|---|---|---|
| 1 | 1 | 3 | 540 | 531.4 | 33.5 | 2.953 | 2.996 | 6.334 | 10÷11 | 16 | 15 | 15 | 18 |
| 1 | 2 | 3 | 540 | 535.2 | 32.9 | 2.966 | 2.913 | 4.158 | 10÷11 | 12 | 15 | 15 | 18 |
| 1 | 3 | 3 | 540 | 534.3 | 33.3 | 3.012 | 2.883 | 6.613 | 10÷11 | 16 | 15 | 15 | 18 |
| 2 | 1 | 1.8 | 200 | 190.5 | 32.4 | 1.842 | 2.971 | 3.845 | 43÷44 | 45 | 49 | 51 | 46 |
| 2 | 2 | 1.8 | 200 | 190.3 | 32.0 | 1.817 | 3.107 | 4.139 | 46÷47 | 48 | 49 | 51 | 46 |
| 2 | 3 | 1.8 | 200 | 190.3 | 32.1 | 1.820 | 3.170 | 3.932 | 47÷48 | 48 | 49 | 51 | 46 |
| 3 | 1 | 3 | 200 | 189.7 | 33.0 | 2.594 | 2.897 | 1.130 | 58÷59 | 43 | 82 | 50 | 70 |
| 3 | 2 | 3 | 200 | 190.1 | 32.6 | 2.578 | 2.939 | 1.270 | 56÷57 | 45 | 82 | 50 | 70 |
| 3 | 3 | 3 | 200 | 189.7 | 32.8 | 2.614 | 2.968 | 1.342 | 57÷58 | 45 | 82 | 50 | 70 |
| 4 | 1 | 4.8 | 200 | 189.8 | 33.7 | 4.899 | 3.004 | 0.049 | 81÷82 | 84 | 100 | 100 | 80 |
| 4 | 2 | 4.8 | 200 | 190.2 | 33.3 | 4.875 | 3.002 | 0.044 | 90÷91 | 84 | 100 | 100 | 80 |
| 4 | 3 | 4.8 | 200 | 190.6 | 33.7 | 4.799 | 3.079 | 0.032 | 88÷89 | 84 | 100 | 100 | 80 |

TABLE 2
CLAMPED-CLAMPED BEAMS: NOMINAL AND MEASURED (*) DIMENSIONS, MEASURED AND CALCULATED PULL-IN VOLTAGES. W.O. STANDS FOR WITHOUT

| Geom. | Sample | t | L | L* | w* | t* | g* | $V_{PI}$* Exper. | $V_{PI}$ w.o. stress Trans126 | Pre-Stress (MPa) | $V_{PI}$ w.o. stress Ansys 2D |
|---|---|---|---|---|---|---|---|---|---|---|---|
| 5 | 1 | 3 | 540 | 541.8 | 32.2 | 2.68 | 2.83 | 57÷58 | 27 | 30 | 29 |
| 5 | 2 | 3 | 540 | 541.0 | 32.3 | 2.7 | 2.81 | 59÷60 | 27 | 32 | 29 |
| 5 | 3 | 3 | 540 | 544.3 | 32.4 | 2.792 | 2.913 | 59÷60 | 29 | 29 | 29 |
| 6 | 1 | 6 | 375 | 371.4 | 13.9 | 5.627 | 3.110 | 180÷190 | 191 | 0 | 195 |
| 7 | 1 | 4.8 | 650 | 650.0 | 11.9 | t*+g*=9.17 | | 88÷89 | 72 | 20 | 70 |
| 7 | 2 | 4.8 | 650 | 653.1 | 11.9 | 6.08 | 3.041 | 88÷89 | 72 | 20 | 70 |
| 7 | 3 | 4.8 | 650 | 655.1 | 12.5 | 6.01 | 3.114 | 88÷89 | 72 | 20 | 70 |

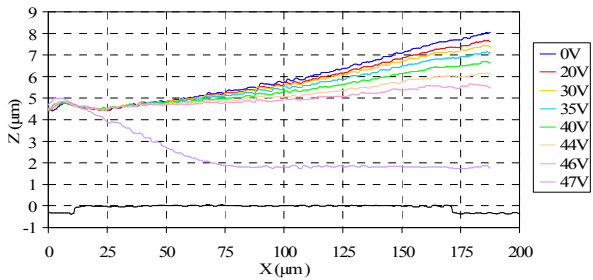

Fig. 2. Sequence of static equilibrium conditions measured by Fogale Zoomsurf 3D during the pull-in tested performed on microcantilever. Geometry 2, sample 2.

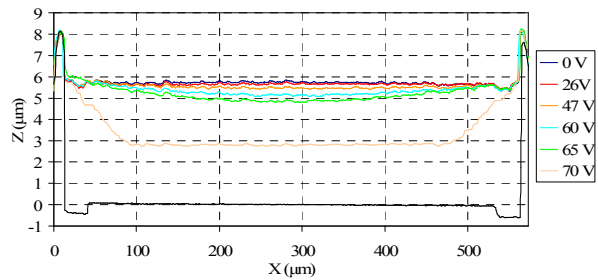

Fig. 3. Sequence of static equilibrium conditions measured by Fogale Zoomsurf 3D during the pull-in tested performed on microbridge. Geometry 5, sample 1.

To predict the actual pull-in of microbridge layout, axial effort had to be identified and included into the FEM models. This investigation was performed in ANSYS by applying either a distributed internal pressure at nodes or an initial strain as real constant. This action allowed estimating residual stress values by tuning the numerical pull-in tension on the experimental result.

IV. DISCUSSION

In Table 1 calculated values of pull-in voltage of microcantilevers are compared to the measured ones. For geometry 1 all methods overestimated pull-in. In particular FEM 3D model was far from the true result. This was due to some problems of mesh morphing in presence of a very narrow gap. Geometry 2 shows a better agreement. Fringing effect is more important and has to be evaluated by a 3D analysis to fit experimental result. In case of geometry 3 a prediction performed by ANSYS worse than by the non incremental approach was observed. Three dimensional effects are more effective in geometry 4, where force inputted in 2D models has to be tuned. Table 2 shows a large difference between pull-in voltage estimated without including residual stress and experimental measures.

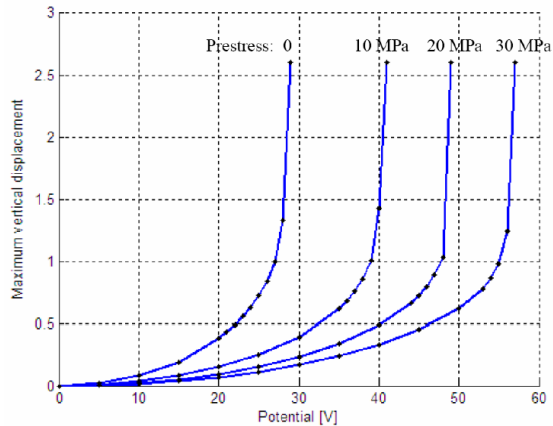

Fig. 4. Analysis of the effect of the tensile load on geometry 5.

Microbridge 5 suffers the highest mismatch. Geometry 6 shows a good agreement and thus a pre-stress almost null. In case of geometry 7 pre-stress was present and detected. Tables demonstrate that upward initial curvature in geometrical nonlinear microcantilever is a very difficult configuration to be analysed. In fact, all 2D and 3D approaches implemented had to operate with a narrow gap and mesh morphing met some problems about pull-in, when the tip is close to the counter-electrode. Reduced order model based on TRANS126 gave better results, since it did mesh dielectric region. Double clamped microbeams strongly suffer pre-stress loading. A sensitivity analysis concerning initial tensile stress was performed. As Fig.4 shows for geometry 5 different pre-stress conditions affect significantly pull-in voltage. It varies with axial stress from $V_{PI}$= 29 (0 pre-stress) to 41 (10 MPa), 49 (20 MPa) and 57 (30 MPa). All models needed to be tuned by inputting a suitable value of tensile stress. In practice, all the microspecimens studied exhibited geometrical nonlinearity, thus requiring to resort to a nonlinear structural analysis based on the iterative solution of the mechanical problem within the sequential approach.

## V. CONCLUSIONS

An experimental investigation was performed to validate numerical approaches aimed to predict the behaviour of microbeams, nonlinearities due to electromechanical coupling and geometry were taken into account. The strong effect of residual stress on the pull-in voltage was detected and included in the models. The experimental results demonstrated a good agreement between the FEM approaches proposed, the methods allow a fairly fast and accurate prediction of the microbeams behaviour.